\documentclass[pra,preprint,groupaddress,floatfix,amsfonts]{revtex4}
\usepackage{amsmath,epsf,dcolumn}
\usepackage{graphicx}
\newcommand{\be}{\begin{equation}}
\newcommand{\ee}{\end{equation}}
\newcommand{\bea}{\begin{eqnarray}}
\newcommand{\eea}{\end{eqnarray}}

\begin{document}

\title{Tightened Lieb-Oxford bound for systems of fixed particle number}

\author{Mariana M. Odashima and K. Capelle} 
\affiliation{Departamento de F\'{\i}sica e Inform\'atica,
Instituto de F\'{\i}sica de S\~ao Carlos,
Universidade de S\~ao Paulo, 
Caixa Postal 369, S\~ao Carlos, 13560-970 SP, Brazil}
\author{S.B.~Trickey}
\affiliation{Quantum Theory Project, Departments of Physics and
of Chemistry, University of Florida, Gainesville, FL 32611, USA}
\date{15 Dec.\ 2008}

\begin{abstract}
The Lieb-Oxford bound is a constraint upon approximate exchange-correlation
functionals.  We explore a non-empirical tightening of that bound in both 
universal and electron-number-dependent form. The test functional is PBE.
Regarding both atomization energies (slightly worsened) and bond
lengths (slightly bettered), we find the PBE functional to be 
remarkably insensitive to the value of 
the Lieb-Oxford bound.  This both rationalizes the use of the original
Lieb-Oxford constant in PBE and suggests that enhancement factors more
sensitive to sharpened constraints await discovery.
\end{abstract}

\maketitle

%

\section{Background}

Construction of approximate exchange-correlation (XC) functionals in
DFT without reliance on empirical data is an
important task, both conceptually and practically.  Perhaps the most
widely used constraint-based approximate XC functional today is the
extremely popular Perdew-Burke-Ernzerhof (PBE) generalized gradient
approximation (GGA)\cite{PBE}.  One of the constraints on which the
PBE GGA X functional (and some others also) is based is the Lieb-Oxford
bound \cite{LO}.  In the DFT literature this bound commonly is
expressed as
\be
\frac{E_{xc}[n]}{E_x^{LDA}[n]} \le \lambda_{LO}
\ee
where
\be
E_x^{LDA}[n]  =  -\frac{3}{4}\left( \frac{3}{\pi} \right)^{1/3} %
\int d^3r n^{4/3}({\mathbf r}).
\label{LObound}
\ee
The LO value for the constant is 
\be
 \lambda_{LO} =  2.273 
\label{LOorig}
\ee
The possibility of tightening this bound has been the
subject of recurrent interest in DFT.  A slightly tighter value
$\lambda_{CH} = 2.215$ was found by Chan and Handy \cite{CH99}.  Vela
\cite{Vela06} later reported that using a spatially varying implementation
of the LO bound which always is tighter than $\lambda_{LO}$ 
improved the results for a test set of light inorganic 
and organic omolecules calculated using 
constraint-based GGAs. 

Shortly thereafter and independently, two of us (OC hereafter)
\cite{ocJCP07,ocIJQC08} gave numerical evidence from exact
and near-exact calculations on atoms, small molecules, and model
systems that the true bound is much tighter.  That analysis
proceeded by defining the \emph{functional} $\lambda[n]$ 
\be
\lambda[n]=\frac{E_{xc}[n]}{E_x^{LDA}[n]} \; ,
\label{lambda-n}
\ee
with both numerator and denominator 
evaluated on the actual density of each system. 
In general, this functional cannot be evaluated 
exactly, because neither $E_{xc}$ nor the density is 
known exactly.  However, it can 
be evaluated to high  accuracy 
for systems for which near-exact XC energies and system 
densities are known from configuration interaction or quantum
monte carlo calculations. The result 
\cite{ocJCP07,ocIJQC08} is that real systems typically have 
$\lambda[n] \in (1.1 \ldots 1.3)$. The  higher end of the interval typically 
corresponds to more rarefied, diffuse density distributions, while 
the lower end corresponds to more compact densities. Values 
above $1.3$ were only found for extreme low-density limits of model 
Hamiltonians: the $k \to 0$ limit of Hooke's atom has $\lambda[k\to 0]=1.489$, 
and the $r_s\to\infty$ limit of the homogeneous electron gas has 
$\lambda[r_s\to\infty]=1.9555=:\lambda_{HEG}$.

On the basis of these results, OC conjectured \cite{ocJCP07,ocIJQC08} that 
further tightening of the LO bound, beyond that obtained by Chan and Handy,
can be achieved, and suggested that for real systems (excluding unphysical 
limits of model Hamiltonians) $\lambda_{OC1}=1.35$ may provide the tightest 
upper limit, whereas for arbitrary systems 
$\lambda_{OC2}=2.00\approx\lambda_{HEG}$ 
is the upper limit. 

OC also speculated that system-specific upper limits could be found,
thereby providing upper limits for all systems sharing some common
properties.  Earlier there was other evidence for system-specific
limits.  Novikov et al. \cite{Novikov97} used a reduced $\kappa$
parameter (defined below) in the PBE XC functional to some benefit.
This reduction (see our discussion below) is equivalent to a reduced
LO bound.  The numerical rationalization for this was published
somewhat later by Peltzer y Blanca et al. \cite{Peltzer01}. Translating
to effective values of $\lambda$, broadly they found that 3d metals do
better with $\lambda \approx \lambda_{LO}$, 4d metals benefit from
$\lambda \approx 1.81 \rightarrow 1.94$, and 5d metals benefit from
$\lambda \approx 1.69 \rightarrow1.84 $.  The notable exception was Fe,
where the effective $\lambda$ was 2.8, an illustration of the fact
that all the limitations of a specified XC form cannot be corrected by
a single parameter fix. (Recently there also has been study of
reduced $\kappa$ in the PBE functional but the reduction is done
in such a way as to respect the original Lieb-Oxford bound \cite{Csonka07},
hence is not directly related to the issue at hand.)

Other than this one empirical example, 
the available data did not allow any general
characterization of $\lambda$-value classes.
Here we propose and explore a generally applicable, 
entirely non-empirical way to characterize classes of systems with a common
maximum value of $\lambda[n]$. This characterization is based on a
rarely mentioned part of the original Lieb-Oxford paper, in which
they show 
that tighter estimates of the upper limit on $\lambda[n]$ can be
achieved by restricting the $\lambda$ functional to densities
which integrate to a specified particle number $N$. We therefore introduce the
\emph{function} $\lambda(N)$, which for a given value of $N$ provides
a universal upper limit upon $\lambda[n]$ valid for all systems 
such that $\int d^3 r n({\mathbf r}) = N$. The maximum value of
$\lambda(N)$, attained for $N\to \infty$, is the value $\lambda_{LO}$
used in common density functionals. The function
$\lambda(N)$ assigns to each class of systems of common particle
number an upper limit $\lambda(N)\leq \lambda_{LO}$.

In construction of constraint-based functionals, 
the fact that the upper limit can be tightened globally (from 
$\lambda_{LO}$ to $\lambda_{CH}$ and perhaps on to $\lambda_{OC2}$) or in a 
system-specific way ({e.g.}, using $\lambda(N)$) has not been taken 
into account, and the consequences of 
a replacement of $\lambda_{LO}$ by one of the lower values in currently 
popular functionals are unknown. We study some of those consequences here.

\section{Construction of a particle-number dependent bound}

To explore the system-specific bound provided by the function
$\lambda (N)$  requires facing the problem that, while Lieb 
and Oxford proved the
existence of this function and deduced some of its properties,  they
did not obtain a closed analytical expression for all $N$. We thus 
propose a simple approximation to $\lambda (N)$, compatible with all
known information on the universal LO bound. The following
facts are known about $\lambda (N)$ \cite{LO,ocJCP07,ocIJQC08}: \\
(i) Its value at $N=1$ is $\lambda(N=1)=1.48 :=\lambda_1$. \\
(ii) Its value at $N=2$ is not known, but is above 
$\lambda_{min}(N=2)=1.67$.\\
(iii) The function $\lambda(N)$ is monotonic, i.e., 
$\lambda(N+1)\ge\lambda(N)$. \\
(iv) Its value at $N=\infty$ is not known, but must be less than or equal to 
$\lambda_{max}(N\to\infty):=\lambda_\infty$. Different proposals for
the value of $\lambda_\infty$ are $\lambda_{LO}=2.273$, $\lambda_{CH}=2.215$,
and $\lambda_{OC2}=2.00\approx\lambda_{HEG}$. \\
(v) The largest value of $\lambda[n]$ found for any system studied
specifically is that for the extreme low-density limit of the
homogeneous electron gas $\lambda_{HEG}(r_s\to\infty)=1.9555$.  For
real physical systems, $\lambda[n]$ typically $\leq 1.3$. These
values provide empirical lower bounds on the function $\lambda(N)$.

Note that standard density functionals either do not make use of the 
Lieb-Oxford bound at all (and some are known to violate it \cite{PW86,BLYP,WL,revpbe})
or exploit only property (iv), normally with the 
weakest value for $\lambda_\infty$, namely $\lambda_{LO}$.  
To construct a model for the function $\lambda(N)$ we exploit properties 
(i) (value at $N=1$), (iii) (monotonicity) and (iv) (value at $N\to\infty$).
We use properties (ii) (theoretical lower limit at $N=2$) and (v) 
($\lambda[n]$ for model and real systems) as consistency tests 
for the construction. With all this in mind, we propose the simple 
interpolation
\be
\lambda(N)=\left( 1 - \frac{1}{N} \right)\lambda_\infty + %
\frac{\lambda_1}{N} \;,
\label{OCNdepLO}
\ee 
where $\lambda_\infty$ is $\lambda_{LO}$, 
$\lambda_{CH}$ or $\lambda_{OC2}$. By construction this interpolation obeys 
properties (i), (iii) and (iv). Direct inspection shows that it also respects 
properties (ii) and (v).

\begin{figure}[t]
\includegraphics[width=15cm]{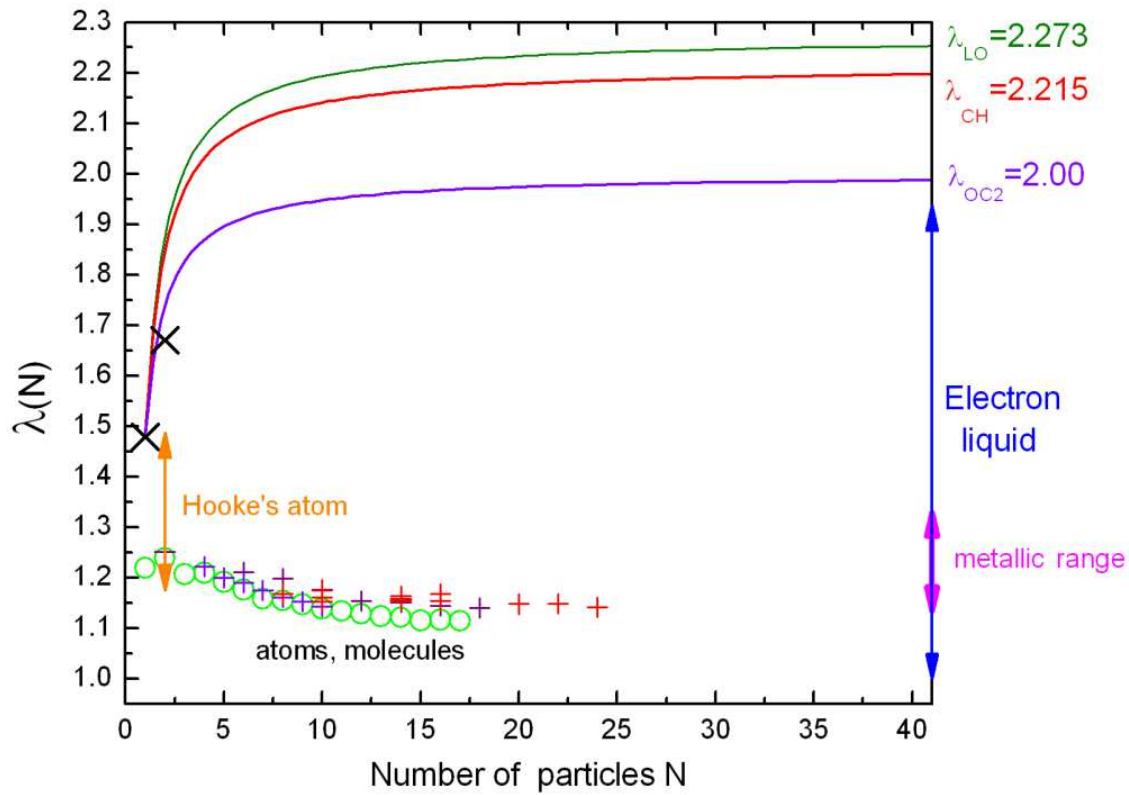}
\caption {\label{fig1} [Color]
Interpolation function $\lambda(N)$ for three different choices of
$\lambda_\infty$, compared to known theoretical results and empirical data.
Black crosses represent the exact value at $N=1$ and the lower bound at $N=2$.
Values at $N\to\infty$ are indicated on the right vertical axis. The three
continuous curves are our interpolation, Eq.\ (\ref{OCNdepLO}), using the three alternative choices
for $\lambda_\infty$. All other data represent ranges or values for selected
real systems, providing empirical lower bounds.}
\end{figure}

Figure~\ref{fig1} illustrates this function for the three different 
choices of $\lambda_\infty$, and compares it to the known value at $N=1$, 
the lower limit at $N=2$, and some representative data for atoms, molecules, 
and the homogeneous electron gas.  

\section{Implementation and Computational Protocols}

\subsection{Modification of PBE GGA} 

To explore these ideas we have implemented the various possible
replacements of $\lambda_{LO}$ in the PBE GGA. At the outset, we
remark that, on the basis of previous experience with the revPBE functional
\cite{revpbe}, we expect that lowering of $\lambda$ in PBE will have a 
detrimental effect on atomic total energies. (In revPBE an {\em increase}
of $\lambda$ was shown to improve atomic total energies and molecular 
atomization energies, at the expense of worsened bond lengths.)

Since the actual values of $\lambda[n]$ for physical systems are known
to fall far below $\lambda_{LO}$, and the theoretical information
available from the CH numerical tightening and from the function
$\lambda(N)$ both indicate that lower values of $\lambda$ are
appropriate, this detrimental effect must be considered a severe
shortcoming of the GGA. An important issue of energetics, therefore,
is whether the atomization energies are improved when tightened LO
bounds are used in a GGA. (Further investigation is needed to see if
meta-GGA functionals \cite{TPSS} suffer from the same problem, but
that is beyond the scope of this study.)

In any event, there are five possibilities for tightening, pertaining
to two categories. Category I is a simple replacement of the constant
value $\lambda_{LO}$ by the alternative lower constants $\lambda_{CH}$
or $\lambda_{OC2}$. Category II replaces the constant by the function
$\lambda(N)$, with the three possible choices for
$\lambda_\infty$. The resulting five choices are to be compared to the
original choice $\lambda_{LO}$, made in the construction of PBE. 

In the PBE GGA, the LO bound is enforced pointwise through the choice
of the parameter $\kappa$ in the exchange enhancement factor
\be
F_{x}^{PBE} := 1 + \kappa - \frac{\kappa}{1 + \mu s^2/\kappa}
\label{origPBEFx}
\ee
with the dimensionless reduced gradient given by 
\be
s({\mathbf r}) = \frac{1}{ 2(3\pi^2)^{1/3}} %
\frac{|\nabla n({\mathbf r})|}{n({\mathbf r})^{4/3}}
\label{sdefn}
\ee
Taking spin-polarization into account, satisfaction of 
Eq.\ (\ref{LObound}) by the enhancement 
factor (\ref{origPBEFx})  is equivalent to 
\be
F_x[n,s] \le \frac{\lambda_{LO}}{2^{1/3}} = 1.804 
\ee
Since $\lim_{s\rightarrow \infty} F_x[n,s] = 1 + \kappa$, the result is 
\be
\kappa_{PBE} = 0.804
\label{PBEkappa}
\ee
Of course, the  simple choice of a different universal bound leads to
\be
\kappa(\lambda_\infty) = \frac{\lambda_\infty}{2^{1/3}} - 1 
\label{kappaUniv}
\ee
The equivalent modification to include the $N$-dependent LO bound,
Eq.\ (\ref{OCNdepLO}),  is 
\be
\kappa(N,\lambda_\infty) = \frac{\lambda(N,\lambda_\infty)}{2^{1/3}} - 1 
\label{kappaN}
\ee
The result of considering such altered LO bounds is five variants of
the PBE X functional:\\ PBEMA: PBE96 X but with $\lambda_\infty =
\lambda_{CH} = 2.215$. \\ PBEMB: PBE96 X but with $\lambda_\infty =
\lambda_{OC2} = 2.00$. \\ PBEMC: PBE96 X but with $\lambda
(N,\lambda_\infty)$ and $\lambda_\infty = \lambda_{LO} = 2.273$. \\
PBEMD: PBE96 X but with $\lambda (N,\lambda_\infty)$ and
$\lambda_\infty = \lambda_{CH} = 2.215$. \\ PBEME: PBE96 X but with
$\lambda (N,\lambda_\infty)$ and $\lambda_\infty = \lambda_{OC2} =
2.00$. \\ The first two, along with the original PBE, 
comprise category I, the latter three,
category II.  We denote the five variants collectively as PBEMx in
what follows.

\subsection{Protocols} 

All five variants were introduced in the code deMon2k, version 2.4.2
\cite{deMon2k}, by systematic modification of the exchange-correlation
modules. Subsequently, the 
implementation was validated by comparison
of atomic calculations done with hard-coded modifications of the code
soatom.f \cite{gtoff}.  Throughout we used the full PBE correlation
functional, not the deMon cutoff version (i.e. we used the deMon2k
``PBESSF'' option), both for ordinary PBE and PBEMx.  
Because deMon2k uses variational Coulomb
fitting, there is a choice of density fitting (auxiliary) basis sets
and of the method for evaluating XC matrix elements.  Initially we
used the so-called A2 density fitting basis (deMon2k option AUXIS(2))
and the option to do the numerical integrals for the XC quantities
using the fitted (auxiliary) density (deMon2k ``AUXIS'' option).  We
return to these options below.

For development of a suitable protocol (Kohn-Sham basis, fitting
basis) we first studied the Li$_2$ molecule in a
triple-zeta-plus-polarization (TZVP) KS basis.
The results are in Table \ref{table1}.  Note that $\Delta E$ is the
total atomization energy, $2\, E_{Li, atom} - E_{Li_2}$ (not
the cohesive energy per atom).
Regarding the quality of the calculation, observe that for the unmodified 
PBE functional, our results are almost identical with those given by 
Ernzerhof  and Scuseria \cite{ES}, $\Delta E = 20$ kcal/mol, 
$R_e = 2.727$\ {\AA}.  (For 
reference, they quote experimental values as $26$ kcal/mol and 
$2.673$\ {\AA}.)

As would be expected from na\"{\i}ve use of  a 
particle-number-dependent model, the results in Table \ref{table1} 
show a clear size-inconsistency problem, signaled by the
big shift in $\Delta E$ between the $N$-independent models, PBE,
PBEMA, PBEMB, and the $N$-dependent models, PBEMC, PBEMD, and PBEME.
The fact that there is no such shift in the $R_e$ values is a clear
sign that the problem is in the comparison with the isolated atom.  
Eq.\ (\ref{kappaN}) illustrates the point.
In a na{\"\i}ve application of 
the $N$-dependent models, the Li atom has $\lambda (3,\lambda_\infty)$
while the Li$_2$ molecule has $\lambda (6,\lambda_\infty)$ (with the
three choices of $\lambda_\infty$). The result is a 
separated atom limit of the diatomic molecule which is not the same as the
isolated atom.  Table \ref{table2} shows the very substantial
difference in the PBE parameter $\kappa$ for these two situations.

\begin{table}[!h]
\caption{
Comparison of effects of various Lieb-Oxford bounds in the PBE X
functional for the Li$_2$ molecule.  See text for notation about 
functionals. $E_{atom}$ and $E_{Li_2}$ are total energies in Hartrees.
$\Delta E$ is the total atomization energy in kcal/mol,  $R_e$ is
the equilibrium bond length in {\AA}.
}
\begin{ruledtabular}
\begin{tabular}{lcccc}
 Functional &  $E_{atom}$ & $E_{Li_2}$ & $\Delta E$ & $R_e$\\
\colrule
PBE    & -7.460992748 & -14.953949056 & 20.06 & 2.7236 \\
PBEMA  & -7.457436406 & -14.946932892 & 20.01 & 2.7218 \\
PBEMB  & -7.441633876 & -14.915640294 & 20.31 & 2.7155 \\
PBEMC  & -7.442385785 & -14.937217688 & 32.91 & 2.7196 \\
PBEMD  & -7.439006265 & -14.930264047 & 32.79 & 2.7181 \\
PBEME  & -7.424518016 & -14.899876726 & 31.90 & 2.7131 \\
 \colrule
\end{tabular}
\end{ruledtabular}
\label{table1}
\end{table}

\begin{table}[!h]
\caption{
Values of the PBE X functional parameter $\kappa(N,\lambda_\infty)$ for
the PBE and PBEMx functionals for $N=3$ (Li atom) and 
$N=6$ (Li$_2$ molecule).
}
\begin{ruledtabular}
\begin{tabular}{lcc}
 Functional &  $\kappa(3,\lambda_\infty)$  &  $\kappa(6,\lambda_\infty)$ \\
\colrule
PBE    & 0.804319  & 0.804319  \\
PBEMA  & 0.757967  & 0.757967  \\ 
PBEMB  & 0.587401  & 0.587401  \\
PBEMC  & 0.594439  & 0.699379  \\
PBEMD  & 0.563537  & 0.660752  \\
PBEME  & 0.449826  & 0.518614  \\
\colrule
\end{tabular}
\end{ruledtabular}
\label{table2}
\end{table}

\begin{table}[!h]
\caption{
Comparison of effects of various Lieb-Oxford bounds in the PBE X
functional for the Li$_2$ molecule. $N$-dependent functionals done
with  $N=6$ size-consistent parameters.  TZVP KS basis. Upper
set is the A2 fitting basis with AUXIS XC evaluation option, 
lower set is GEN-A2 and BASIS option. See text for details as
well as notation for functionals. 
$E_{atom}$ and $E_{Li_2}$ are total energies in Hartrees. $\Delta E$ 
is the total atomization energy in kcal/mol. $R_e$ is
the equilibrium bond length in {\AA}.
}
\begin{ruledtabular}
\begin{tabular}{lcccc}
 Functional &  $E_{atom}$ & $E_{Li_2}$ & $\Delta E$ & $R_e$\\
\colrule
PBE    & -7.460992748 & -14.953949056 & 20.06 & 2.7236 \\
PBEMA  & -7.457436406 & -14.946932892 & 20.01 & 2.7218 \\
PBEMB  & -7.441633876 & -14.915640294 & 20.31 & 2.7155 \\
PBEMC  & -7.452520577 & -14.937217688 & 20.19 & 2.7196 \\
PBEMD  & -7.449007723 & -14.930264047 & 20.24 & 2.7181 \\
PBEME  & -7.433703689 & -14.899876726 & 20.37 & 2.7131 \\
 \colrule
PBE    & -7.460613173 & -14.953310471 & 20.13 & 2.7304 \\
PBEMA  & -7.457076546 & -14.946319641 & 20.18 & 2.7277 \\
PBEMB  & -7.441303732 & -14.915029100 & 20.34 & 2.7181 \\
PBEMC  & -7.452179554 & -14.936623118 & 20.25 & 2.7244 \\
PBEMD  & -7.448674591 & -14.929672255 & 20.28 & 2.7222 \\
PBEME  & -7.433359029 & -14.899212818 & 20.39 & 2.7143 \\
 \colrule
Reference values \cite{Eoatoms,Eomols,ExpAtEn,ExpBondLen}  & -7.47806     & -14.9938      & 26    & 2.673  \\
 \colrule
\end{tabular}
\end{ruledtabular}
\label{table3}
\end{table}

Table \ref{table3} shows how a size-consistent set of parameters, here
for $N=6$, resolves the problem.  (For clarity, note we made
the common choice throughout all these calculations and ignored the
DFT spin-symmetry problem. Thus, the separated atoms are
spin-polarized even though the molecule has multiplicity equal one.)
Throughout this study, we used this same  size-consistent 
procedure, namely applying to the separated atoms the modified
LO constants proper for the value of $N$ of the aggregated 
system (molecule) in question.  For
heteronuclear molecules, especially hydrides, this protocol results
in a rather disparate enforcement of the LO bound for atoms of
substantially different $N$, a matter for later study and refinement.
(We note that the use of the original PBE functional implies the
most disparate enforcement of all, as it amounts to using the
largest $N\to\infty$  value of $\lambda$ for all finite $N$.)

Table \ref{table3} also compares the effect of the two different
options in deMon2k for evaluation of the XC matrix elements. First is
the deMon A2 density fitting basis (deMon2k option AUXIS(2)) and the
aforementioned deMon2k option (``AUXIS'') for evaluation of XC
quantities using the fitted (auxiliary) density on a numerical grid.
Second is the richer GEN-A2 fitting basis and evaluation of the the XC
quantities from the density formed straightforwardly from the KS
orbitals also on the numerical grid (``BASIS'' option).  In principle,
the latter procedure is the more accurate and is the one we adopted.
Nevertheless, the trends in the PBEMx series are
essentially the same in the  less-accurate procedure.

Our other exploratory test was the O$_2$ molecule, a triplet ground
state system.  The TZVP PBE atomization energy (see the first two
lines of Table \ref{table4}) is about three percent off from the
published result of Ernzerhof and Scuseria \cite{ES}, who used the
substantially richer basis 6311+G(3df,2p).  An ACES-II \cite{ACES-II}
calculation using another rich basis (aug-cc-PVTZ) matched the deMon2k
results with that same KS basis and the richer density-fitting basis
(``GEN-A2*'' option).  These results, in the third and fourth lines of
Table \ref{table4}, calibrate the effects of basis set differences.
Results for the PBEMx series in the aug-cc-PVTZ/GEN-A2* basis sets
also are in Table \ref{table5}.  The relative shifts among the six PBE
variants are the same irrespective of basis sets, but the richer basis
sets make the atomization energies larger and bond lengths slightly
shorter.  Ernzerhof and Scuseria \cite{ES} quote the experimental data
as $\Delta E$ = 118 kcal/mol and $R_e$ = 1.208 {\AA}.  Thus, all six
PBE variants (original plus five new give too deep a binding energy at
slightly elongated bond lengths.

\begin{table}[!h]
\caption{ 
O and O$_2$ (triplet) total energies ($E_H$), molecular
atomization energy (kcal/mol), and equilibrium bond length ({\AA}) 
for the ordinary PBE XC functional as
calculated in deMon2k, TZVP basis (``deMon-TZVP''); Gaussian03, 
6311+G(3df,2p) basis  (Ref.\ \cite{ES} ``ES''); 
 ACES-II, aug-cc-PVTZ (Ref.\ \cite{AP} ``AP''); deMon2k, aug-cc-PVTZ
basis (``deMon-aug'').
}
\begin{ruledtabular}
\begin{tabular}{lcccc}
Calc. &  $E_{tot,O}$ & $E_{tot,O_2}$ & $\Delta E$ & $R_e$\\
\colrule
deMon-TZVP  & -75.00612438 & -150.23282532 & 138.4 & 1.23491 \\
ES    &  & & 143 & 1.217 \\
AP    & -75.00773627 & -150.24372619 & 143.2  & 1.21996 \\
deMon-aug  & - 75.00781596 & -150.24387554 & 143.2  & 1.22008 \\
 \colrule
\end{tabular}
\end{ruledtabular}
\label{table4}
\end{table}

\begin{table}[!h]
\caption{
As in Table \ref{table3} but for triplet O$_2$ and for two different
basis sets TZVP/GEN-A2 (upper set), aug-cc-PVTZ/GEN-A2* (lower set).
}
\begin{ruledtabular}
\begin{tabular}{lcccc}
 Functional &  $E_{atom}$ & $E_{O_2}$ & $\Delta E$ & $R_e$\\
\colrule
PBE    & -75.00612438 & -150.23282532 & 138.4  & 1.2349 \\
PBEMA  & -74.99689680 & -150.21624772 & 139.6  & 1.2343 \\
PBEMB  & -74.95432779 & -150.13915494 & 144.6  & 1.2315 \\
PBEMC  & -74.99835094 & -150.21886346 & 139.4  & 1.2344 \\
PBEMD  & -74.98900193 & -150.20202494 & 140.6  & 1.2338 \\
PBEME  & -74.94639261 & -150.12467907 & 145.5  & 1.2310 \\
 \colrule
PBE    & -75.00781596 & -150.24387554 & 143.2  & 1.2201 \\
PBEMA  & -74.99853463 & -150.22722985 & 144.4  & 1.2195 \\
PBEMB  & -74.95571887 & -150.14986185 & 149.6  & 1.2167 \\
PBEMC  & -74.99998015 & -150.22985618 & 144.3  & 1.2196 \\
PBEMD  & -74.99059055 & -150.21295118 & 145.4  & 1.2189 \\
PBEME  & -74.94774275 & -150.13534295 & 150.5  & 1.2162 \\
 \colrule
Reference values \cite{Eoatoms,Eomols,ExpAtEn,ExpBondLen}  & -75.0674     & -150.2770     & 118    & 1.208  \\
 \colrule 
\end{tabular}
\end{ruledtabular}
\label{table5}
\end{table}

The preceding discussion makes clear that systematic comparison
of the five PBE variants generally does not require a fully saturated
basis set. Rare gas dimers, discussed below, are an exception.
Therefore, except for rare gas dimers, we adopted the following
protocol: (i) use TZVP for the KS basis; (ii) Use GEN-A2 or GEN-A3
algorithms to generate the fitting-function basis (and thereby
minimize the effect of the variational Coulomb fitting, which gives a
lower bound to the Coulomb repulsion that can be deceptive with a
poorly chosen fitting basis); (iii) use the deMon2k BASIS option for
evaluation of the XC matrix elements. This protocol combines a
reasonably rich KS basis with an abundance of caution in treating the
XC quantities.

For the rare gas dimers, test calculations on Ar$_2$ with both a DZVP
and a 6-311++G(3df,3pd) KS basis set demonstrated that these do not
reproduce known, large-basis PBE results for this dimer
\cite{Patton97,Zhao06}.  Since those two calculations were completely
independent and gave essentially identical values, $\Delta E = 0.138$
kcal/mol, $R_e = 4.00$\ \AA, it is essential to reproduce them.
Therefore, we shifted to the aug-cc-pVTZ KS basis \cite{EMSL}, as used
by Zhao and Truhlar, and the deMon2k GEN-A3 fitting function basis.
This combination gives the same PBE results as the foregoing two
references. We treated Ne$_2$ with the corresponding aug-cc-pVTZ KS
basis and GEN-A3 fitting function basis.

\section{Results and Discussion}

\begin{figure}      
\epsfxsize=7.20cm
\vspace{0.5cm}
\centerline{\hspace*{+15pt}\epsffile{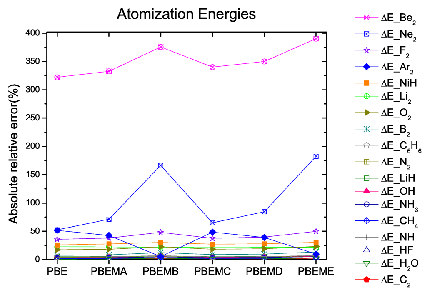}}
\vspace{1.1cm}
\epsfxsize=7.20cm
\centerline{\hspace*{+15pt}\epsffile{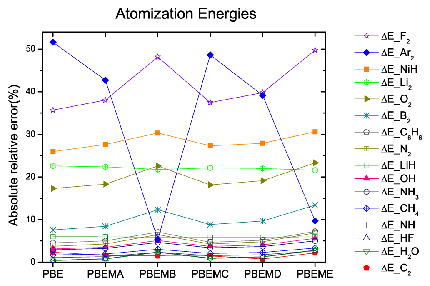}}
\vspace{1.1cm}
\epsfxsize=7.20cm
\centerline{\hspace*{+15pt}\epsffile{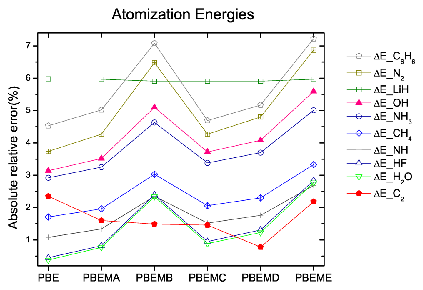}}
\vspace{0.3cm}
\caption{
(Top panel): Absolute relative errors in atomization energies for all
18 molecules for original PBE and the five variants.  
(Middle panel): As in the upper panel but with the worst two cases
(Be$_2$, Ne$_2$) removed to allow a finer scale. 
(Bottom panel): As in the middle panel but with the worst six
cases of that panel (F$_2$, Ar$_2$, NiH, Li$_2$, O$_2$, B$_2$) removed to allow a finer scale.
}
\label{fig2}
\end{figure}

\begin{figure}      
\epsfxsize=7.20cm
\vspace{0.5cm}
\centerline{\hspace*{+15pt}\epsffile{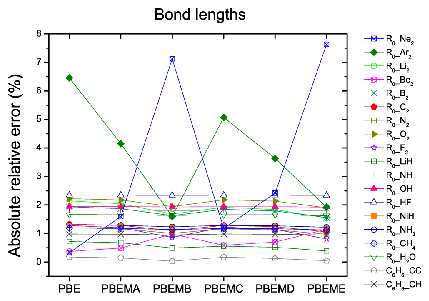}}
\vspace{1.1cm}
\epsfxsize=7.20cm
\centerline{\hspace*{+15pt}\epsffile{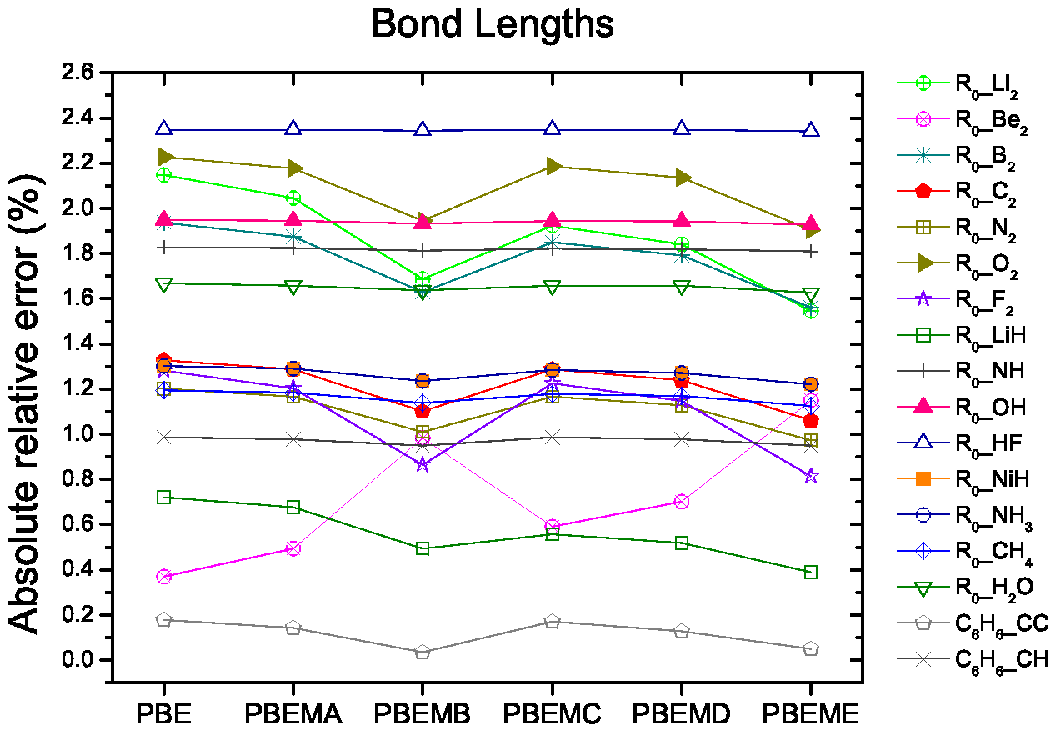}}
\vspace{0.3cm}
\caption{
(Upper panel): Absolute relative errors in bond lengths for all
18 molecules for original PBE and the five variants.  
(Lower panel): As in the upper panel but with the two most sensitive cases
(Ar$_2$, Ne$_2$) removed to allow a finer scale. 
}
\label{fig3}
\end{figure}

For PBE and PBEMx, Table \ref{table6} gives the atomization energies
for eighteen light molecules, while \ref{table7} gives bond lengths
and bond angles for those same systems.   (As a technical 
aside, note that NiH is a difficult system to treat.)  Absolute
relative errors in the atomization energy  are shown  in 
Figure \ref{fig2} and the corresponding bond length data are in 
Figure \ref{fig3}.

Several features stand out from these results.  With a few exceptions,
the general pattern is that both atomization energies and bond lengths
are remarkably insensitive to changes in the enforcement of the LO
bound.  This outcome is consistent with what one might have intuited
from Figure~\ref{fig1}.  The $\lambda(N)$ which we are justified 
in using (in the sense that our interpolation respects the known
constraints) is substantially larger than the $\lambda$ values
imputed for actual molecules.  So one might conclude that a more
refined way of implementing the LO bound in a GGA is needed.

However, two systems, Ne$_2$ and Ar$_2$, are notably sensitive 
to the value of $\lambda$ in the atomization energies. A
coherent interpretation by classes of molecules is possible:
unlike the other atoms, Ne and Ar are closed shell systems. Thus, we may 
suspect that the well-known peculiarities of closed shell
interactions are the source of the distinct behavior.  For the 
equilibrium bond lengths Ne$_2$ and Ar$_2$ still stand out from 
all other systems by being most sensitive to
changes in $\lambda$. (See Fig.\ \ref{fig3}.) Interestingly,
the Be$_2$  bond-length variation is at odds with all the other 
open-shell systems. The 
behavior of Be$_2$  does not seem to be traceable to being from the rapidly
varying part of the $\lambda(N)$ function, since Li$_2$ is in that
region also and it is insensitive in both bond length and atomization
energy.  


To display the effects of imposition of the size-consistent molecular
$N$ values on the LO bounds in atoms, we also calculated some isolated
atom total energies  at their intrinsic $N = Z$ values.
For H, C, N, O, and F, Table \ref{table8} displays the results for the
intrinsic value versus the results for the highest $N$ molecule in
which each element was used in the present study.  As would be
expected from the interpolation, Eq.\ (\ref{OCNdepLO}), and the
constraints on which it is based, the total energy of the H atom
exhibits the largest percentage variation between intrinsic and
molecular values for for the three $N$-dependent variants (PBEMC,
PBEMD, PBEME).

\begin{table}[!h]
\caption{
Comparison of effect 
various Lieb-Oxford bounds in the PBE X
functional upon the atomization energies (kcal/mol) of 
various small molecules.  See text for notation about 
functionals. 
}
\begin{ruledtabular}
\begin{tabular}{lcccccccc}
Species & $2S+1$ & PBE & PBEMA & PBEMB & PBEMC & PBEMD & PBEME & Exp \cite{ExpAtEn}\\
\colrule
Li$_2$ & 1 & 20.13 & 20.18 & 20.34 & 20.25 & 20.28 & 20.39 & 26 \\
Be$_2$ & 1 & 9.71 & 9.95 & 10.94 & 10.12 & 10.35 & 11.28 & 2.3\\
B$_2$ & 3 &  76.7 & 77.3 & 80.1 & 77.6 &  78.2 & 80.8 & 71.3\\
C$_2$ & 1 & 142.3 & 143.4 & 147.9 & 143.6 & 144.6 & 148.9 & 146\\
N$_2$ & 1 & 235.5 & 236.7 & 241.7 & 236.7 & 237.9 & 242.6 & 227 \\
O$_2$ & 3 & 138.4 & 139.6 & 144.6 & 139.4 & 140.6 & 145.5 & 118\\
F$_2$ & 1 & 51.56 & 52.46 & 56.30 & 52.23 & 53.12 & 56.90 & 38\\
Ne$_2$ & 1 &0.1279 &0.1438 &0.2236 &0.1385 & 0.1549 & 0.2369 & 0.0839\\
Ar$_2$ & 1 &0.1377 &0.1631 &0.2998 &0.1462 &0.1732& 0.3122 & 0.2846 \\
HF  & 1 & 142.645 & 143.161 & 145.401 & 143.356 & 143.864 & 146.037 & 142 \\
LiH & 1 & 54.53 & 54.54 & 54.57 & 54.57 & 54.57 & 54.54 & 58 \\
OH  & 2 & 110.361 & 110.762 & 112.458 & 110.976 & 111.365 & 112.981 & 107\\
NH  & 3 & 88.95 & 89.18 & 90.07 & 89.34 & 89.55 & 90.36 & 88 \\
NiH & 2 & 74.05 & 75.05 & 76.63 & 74.86 & 75.21 & 76.79 & 58.8\\
H$_2$O & 1 & 235.9 & 236.8 & 240.5 & 237.1 & 237.9 & 241.5 & 235 \\
NH$_3$ & 1 & 305.686 & 306.664 & 310.761 & 307.029 & 307.976 & 311.882 & 297\\
CH$_4$ & 1 & 427.184 & 428.245 & 432.736 & 428.641 & 429.673 & 433.983 & 420 \\
C$_6$H$_6$ & 1 & 1423.74 & 1430.39 & 1458.48 & 1425.84 & 1432.45 & 1460.30 & 1362 \\
 \colrule
\end{tabular}
\end{ruledtabular}
\label{table6}
\end{table}

\begin{table}[!h]
\caption{
Comparison of effect 
of various Lieb-Oxford bounds in the PBE X
functional upon the bond lengths  (\AA ) and bond angles (degrees)
of various small molecules. The NH$_3$ bond angle is $\theta_{HNH}$ 
See text for notation about  functionals.  CH$_4$ was done with
$T_d$ symmetry enforced.
}
\begin{ruledtabular}
\begin{tabular}{lccccccc}
Species &  PBE & PBEMA & PBEMB & PBEMC & PBEMD & PBEME & Exp \cite{ExpBondLen} \\
\colrule
Li$_2$ & 2.7304 & 2.7277 & 2.7181 & 2.7244 & 2.7222 & 2.7143 & 2.673\\
Be$_2$ & 2.4409 & 2.4379 & 2.4259 & 2.4355 & 2.4328 & 2.4218 & 2.45\\
B$_2$ & 1.6208 & 1.6198 & 1.6159 & 1.6194 & 1.6185 & 1.6148 & 1.590\\
C$_2$ & 1.2595 & 1.2590 & 1.2567 & 1.2590 & 1.2584 & 1.2562 & 1.243\\
N$_2$ & 1.1112 & 1.1108 & 1.1091 & 1.1108 & 1.1104 & 1.1087 & 1.098\\
O$_2$ & 1.2349 & 1.2343 & 1.2315 & 1.2344 & 1.2338 & 1.2310 & 1.208\\
F$_2$ & 1.4301 & 1.4290 & 1.4242 & 1.4293 & 1.4282 & 1.4235 & 1.412\\
Ne$_2$ & 3.0808 & 3.0418 & 2.8709 & 3.0546 & 3.0162 & 2.8550 & 3.091\\
Ar$_2$ & 3.99907 & 3.9124 & 3.6964 & 3.9469 & 3.8929 & 3.6841 & 3.7565\\
HF  & 0.9385 & 0.9385 & 0.9385 & 09385 & 0.9385 & 0.9385 & 0.917\\
LiH & 1.6065 & 1.6058 & 1.6029 & 1.6039 & 1.6033 & 1.6012 & 1.595\\
OH  & 0.9899 & 0.9899 & 0.9898 & 0.9899 & 0.9899 & 0.9897 & 0.971\\
NH  & 1.0549 & 1.0549 & 1.0548 & 1.0549 & 1.0549 &1.0548 & 1.036\\
NiH & 1.4580 & 1.4594 & 1.4553 & 1.4599 & 1.4590 & 1.4549 & 1.477\\
H$_2$O, R & 0.9750  &0.9749 & 0.9747& 0.9749 & 0.9749& 0.9746 & 0.959\\
H$_2$O, $\theta$ &104.21 &104.24  & 104.39 & 104.25 & 104.29 & 104.43 & 103.9\\
NH$_3$, R & 1.0252 & 1.0252 & 1.0245 & 1.0250 & 1.0249 & 1.0244 & 1.012\\
NH$_3$, $\theta$ & 106.40 & 106.44 & 106.62 & 106.45 & 106.49 & 106.67 & 106.7\\
CH$_4$ & 1.0990 & 1.0989 &1.0984 & 1.0988 & 1.0987 & 1.0982 & 1.086\\
C$_6$H$_6$, R$_{CC}$ & 1.3995 & 1.3990 & 1.3965 & 1.3994 & 1.3988 & 1.3963 & 1.397\\
C$_6$H$_6$, R$_{CH}$ & 1.0947 & 1.0946 & 1.0943 & 1.0947 & 1.0946 & 1.0943 & 1.084\\
 \colrule
\end{tabular}
\end{ruledtabular}
\label{table7}
\end{table}

\begin{table*}[!h]
\caption{
Total energies (Hartree a.u.) for five chemically important %
atoms from various Lieb-Oxford bounds %
in the PBE X functional. Results for the N-dependent functionals are %
given both for the values of $N$ intrinsic to the specific atom %
and for the highest molecular $N$ used: 42 for H, 42 for C, 14 for N, %
16 for O, 18 for F.}
\begin{ruledtabular}
\begin{tabular}{lccccc}
XC &  H &  C & N & O & F  \\
\colrule
PBE & -0.498147969 & -37.794851185 & -54.530389203 & -75.006124382 & %
 -99.664580137 \\
PBEMA & -0.497515476 & -37.787528107 & -54.522145702 &  -74.996896796 & %
 -99.654626939 \\
PBEMB & -0.494726517 & -37.754129975  & -54.484419695 & -74.954327792 & %
 -99.608420218 \\
PBEMC & -0.481206885 & -37.777286824 & -54.513676785 & -74.989918352 & %
 -99.649171204 \\
(intrinsic $N$) &&&&&\\
PBEMC  & -0.497950959 & -37.792580522  & -54..52240237 & -74.99835094  & %
 -99.65716700 \\
(highest $N$) &&&&&\\
PBEMD &-0.481206885 & -37.769885027 & -54.505322010 & -74.980521867 & %
 -99.639010126 \\
(intrinsic $N$) &&&&&\\
PBEMD  & -0.497316032 & -37.785199634 & -54.51406717 &-74.98900193  & %
 -99.64707222   \\
(highest $N$) &&&&&\\
PBEME &-0.481206885 & -37.736869483 & -54.467831152 & -74.937949507 & %
 -99.592621018 \\
(intrinsic $N$) &&&&&\\
PBEME  & -0.494539297 & -37.751826019 &-54.47640949  & -74.94639261 & %
 -99.60074422 \\
(highest $N$) &&&&& \\
 \colrule
 exact \cite{Eoatoms} & -0.5 & -37.8450 & -54.5893 & -75.0674 & -99.7341 \\
 \colrule
\end{tabular}
\end{ruledtabular}
\label{table8}
\end{table*}

\section{Concluding Remarks}

Our results show that PBE is rather insensitive to changes in $\lambda$ for
atoms and covalently and ionically bound small molecules. Overall, a reduced, and thus,
in principle, better, value of $\lambda$ produces slightly worsened
energies and slightly improved bond lengths.  This insensitivity
explains why PBE can be successful even though it uses the
$\lambda_\infty$ value even for small $N$.  In this sense, the present
study provides additional insight into the success of PBE for small
systems. On the other hand, a suitably designed, constraint-based
functional should give improved results when the constraints it incorporates
are sharpened.  The failure of PBE to meet this expectation must be
considered a limitation of the PBE functional form.

In the case of the closed shell systems, we find a more pronounced
 $\lambda$ dependence than in the covalent and ionic systems.
Because of the delicate nature of binding in these 
systems, more detailed investigation would be needed to make 
conclusive statements.  In the spirit of the preceding
paragraph, it would appear to be more productive to focus on 
developing enhancement factors that are more sensitive to
sharpening of constraints.

An appealing thought is that the insensitivity found here may 
also have to do with the way
that the LO bound is implemented in DFT in general.  The original
LO bound is for the Coulomb exchange and correlation energy $W_{xc}$ and
does not include the correlation kinetic energy, $T_c = T - T_s \ge 0$, which 
contributes to $E_{xc}$.  As a result, $E_{xc} \ge W_{xc}$ and the
functional $\lambda[n]=E_{xc}[n]/E_x^{LDA}[n]$ which was evaluated in 
\cite{ocJCP07,ocIJQC08} is smaller than the functional
\be
\lambda_W[n] := \frac{W_{xc}[n]}{E_{x}^{LDA}[n]} \;.
\ee
If the effect of T$_c$ were large enough, it might explain at least part
of the large difference
between the values of $\lambda[n]$ and $\lambda(N)$ in Figure \ref{fig1}.
What limited numerical evidence we have, however, suggests that
$\lambda_W[n]$ is only about 10\% larger than $\lambda[n]$, 
a very modest shift compared to the difference in Figure \ref{fig1}.

The more general point, however, that the LO bound is a constraint
on  exchange \emph{and} correlation together, seems to be sustained 
by our findings, in that the PBE form enforces the bound purely
on exchange.  One speculation is that the insensitivity found here
is in part a consequence of that restricted use of the LO bound.

Finally, we consider aspects of $N$-dependence and chemical
classification for enforcement of the LO bound. The first insight is
that, in retrospect, $N$-dependent satisfaction of the LO bound
actually arose very early in DFT, before the LO proof.  In Slater's
X$\alpha$ model, $E_{xc}$ is modeled by scaling $E_{x}^{LDA}$.  (From
a modern perspective, X$\alpha$ is a one-parameter XC model which
gains simplicity at the cost of violating correct scaling for C.)  The
$\alpha$ parameter is $N$-dependent \cite{Slater} and exhibits very
clear shell structure \cite{Schwarz72,Schwarz74}.  For X$\alpha$, the
LO functional $\lambda [n]$ of Eq.\ (\ref{lambda-n}) 
is just $3 \alpha / 2$.
With typical values of $\alpha$ \cite{Schwarz72,Schwarz74}, this gives
$\lambda[n] = 1.0745$ for H y($N=1$) to $1.0387$ for Rn ($N=86$).
Comparison with Fig.\ \ref{fig1} shows that these values are slightly
smaller than the highly accurate empirical values found in Refs.\
\cite{ocJCP07,ocIJQC08}.  

The importance of shell-dependent classification was evident in the
modern work of Refs.\ \cite{ocJCP07} and \cite{ocIJQC08}. The
numerical results of this study also leave a strong suggestion that
such classification would be helpful.  An advantage of the present
classification of $\lambda$ with respect to $N$ alone is that it can
be done in an entirely non-empirical way, as it relies only upon exact
properties of the function $\lambda(N)$. (This is a clear distinction
from all parameterized approaches.)  Of course, the choice of
interpolating function is not unique, but the fact that the upper
limit on $\lambda$ depends on $N$ is completely general. So are all
the properties of $\lambda(N)$ used in the construction of our
interpolation. What this means is that whatever shell-dependent
classification might be invented, it must somehow be an addition to
(or incorporate) classification by particle number, not supplant it.
Because that classification will have to avoid size-inconsistency, we
suspect that the formulation will require additional insight,
including additional constraints.

\section{Acknowledgments}

SBT thanks  Ajith Perera for the ACES-II calculations on O$_2$,
and  Andreas K\"oster, Gerald Geudtner, and Patrizia
Calaminici (Cinvestav, M\'exico DF)  for technical advice on 
the use of deMon2k. MMO was supported by FAPESP.  
KC was supported by FAPESP and CNPq. SBT was supported in part by the
U.S.~National Science Foundation under DMR-0325553 (ITR).

\end{document}